\documentclass{ws-procs975x65}

\usepackage{pstcol}
\usepackage{pst-node}
\usepackage[all]{xy}
\usepackage[innercaption]{sidecap}
\usepackage{wrapfig}
\usepackage{colortbl}

\definecolor{light-yellow}{rgb}{1,1,0.784}
\definecolor{light-blue}{rgb}{0.8,0.85,1}
\definecolor{light-red}{rgb}{0.98,0.933,1}
\definecolor{llight-red}{rgb}{0.98,0.933,1}

\newpsobject{showgrid}{psgrid}{subgriddiv=1,griddots=10,gridlabels=6pt}

\newcommand{\Show}[2]{\psshadowbox[fillstyle=solid,fillcolor=#1]{\txt{#2}}}

\newcommand{\ShowX}[3]{{\Show{#1}{\begin{minipage}[t]{#2}\center{#3}\end{minipage}}}}

\newcommand{\ShowL}[3]{{\Show{#1}{\begin{minipage}[t]{#2}{#3}\end{minipage}}}}

\begin{document}

\title{General Relativity in Space and Sensitive Tests of the Equivalence Principle}

\author{C. L\"AMMERZAHL}

\address{${}^1$ ZARM, University of Bremen, Am Fallturm, 28359 Bremen, Germany \\
E-mail: laemmerzahl@zarm.uni-bremen.de}  

\maketitle

\abstracts{
An introduction into the fundamental quests addressed in space missions is given. These quests are the exploration of the relativistic gravitational field, the Universality of Free Fall, the Universality of the Gravitational Redshift, Local Lorentz Invariance, the validity of Einstein's field equations, etc. In each case, the correspoding missions take advantage of the space conditions which are essential for the improvement of the accuracy of the experiments as compared to experiments on ground. A list and a short description of past, current and planned projects is given. Also the key technologies employed in  space missions are addressed.
}

\section{Introduction}

Fundamental Physics is becoming very exciting these days. The reasons for that are twofold: On the theoretical side the unification of General Relativity and quantum theory seems to lead to deviations from standard physics. And on the experimental side new developments of high precision apparatus make new realms of physics accessible leading to new tests and to new observations. Consequently, the expectation for ''new physics'' as well as improvements in experimental devices strongly push the efforts for the realization of new experiments and observations. An important aspect in this connection is the quality of the experimental environment. It is clear that most experiments need a noise--free environment with stable thermal, seismic, electric etc. conditions. Furthermore, some experiments might profit a lot if they will be carried out in a free--fall and non--rotating environment and some experiments necessarily require that environment. This leads one to the conclusion that there are quite a few experiments which -- when carried through in space on a satellite or on the ISS -- will give results which are orders of magnitude better than when carried through on Earth. 

\begin{figure}[t]
\psset{unit=1.1cm}
\begin{center}
\begin{pspicture}(-3,-2)(3,3)
\pscircle[fillstyle=vlines,hatchcolor=light-yellow,hatchangle=0](0,1){2}
\pscircle[fillstyle=hlines,hatchcolor=light-blue,hatchangle=-30](0.866025,-0.5){2}
\pscircle[fillstyle=hlines,hatchcolor=light-red,hatchangle=210](-0.866025,-0.5){2}
\rput(0,2){\txt\footnotesize\sffamily{quests in \\ fundamental \\ physics}}
\rput(1.93205,-1){\txt\footnotesize\sffamily{experimental \\ develop- \\ ments}}
\rput(-1.93205,-1){\txt\footnotesize\sffamily{advantages \\ of space- \\ conditions}}
\rput(0,-0.2){\txt\footnotesize\sffamily{fundamental- \\ physics \\ experiments \\ in space}}
\end{pspicture}
\end{center}
\medskip
\caption{The conditions for space projects.\label{Conditions}}
\end{figure}
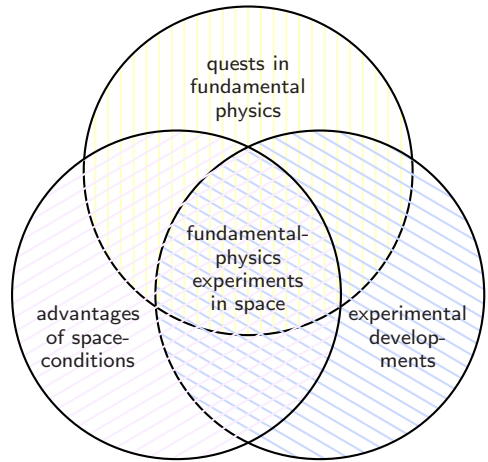

In the following we will shortly review these three aspects, namely the status of the fundamental quests, new experimental developments and the conditions in space. In the case that these three conditions complement one another appropriately, it is reasonable to think about doing these experiments in space, see Fig.\ref{Conditions}. 

The first dedicated space mission for fundamental physics was GP-A which measured the gravitational redshift with an until today unrivalled accuracy. It was during the last years that it was recognized that for many experiments space conditions are really indispensable. As a consequence, many space missions have been proposed. Though most of these proposed missions are definitely worth to be carried through, the huge expenses, long planning time and troublesome efforts for space qualification of experimental devices cut down the number of missions expected to fly to a very few. This is a big disadvantage which might be overcome by developing and using more standardized space techniques. Furthermore, in some cases it is more efficient, instead of developing a dedicated mission, to make use of the ISS which is already existing -- even at the price of experimental conditions which are not optimal (see Sec.4). 

For the convenience of the reader we give a list of acronyms of the missions mentioned in this short review:

\begin{tabbing} 
MICROSCOPE \= Erklaerung  \kill 
ACES \> Atomic Clock Ensemble in Space \\
ASTROD \> Astrodynamical Space Test of Relativity using Optical Devices \\
BEST \> Boundary Effects near Superfluid Transitions \\
DYNAMX \> critical dynamics in microgravity \\
EXACT \> EXperiments Along Coexistence near criTicallity \\
GG \> Galileo Galilei \\
GP-A \> Gravity Probe A \\
GP-B \> Gravity Probe B \\
HYPER \> HYPER precision atomic interferometer in space \\
LAGEOS \> LAser GEOdynamic Satellite \\
LATOR \> Laser Astrometric Test Of Relativity \\
LISA \> Laser Interferometer Space Antenna \\
LLR \> Lunar Laser Ranging \\
MICROSCOPE \> MICRO--Satellite a train\'ee Compens\'ee pour l'Observation du Principe \\ 
\> d'Equivalence \\ 
MISTE \> MIcrogravity Scaling Theory Experiment \\
OPTIS \> OPtical Test of the Isotropy of Space \\
PHARAO \> Projet d'Horloge Atomique par Refroidissement d'Atomes en Orbite \\
PARCS \> Primary Atomic Reference Clock in Space \\
RACE \> Rubidium Atomic Clock Experiment \\
SEE \> Satellite Energy Exchange \\
STEP \> Satellite Test of the Equivalence Principle \\
STM \> SpaceTime Mission \\
SUE \> Superfluid Universality Experiment \\
SUMO \> SUperconducting Microwave Oscillator \\
\end{tabbing} 

\section{Fundamental physics}

\subsection{The general scheme}

Today's fundamental physics is characterized by two schemes\cite{LaemmerzahlDittus02}, see Table \ref{Theories}: By the universal theories and by the four interactions. The universal theories which are applicable to any kind of physical phenomenon are (i) quantum theory, (ii) Special Relativity (SR), (iii) General Relativity (GR), and (iv) many particle physics. The interactions are (i) gravity, (ii) electromagnetism, (iii) the weak, and (iv) the strong interaction. Exept for gravity, they have been successfully unified. It can be seen that gravity is exceptional since it appears on both sides: it is universal due to its universal couplings and it is a particular interaction. 

\begin{table}[b]
\tbl{Universal theories and the four presently known interactions.}
{\sffamily\footnotesize
\begin{tabular}{|>{\columncolor{white}}p{5.2cm}|>{\columncolor{white}}p{5.2cm}|} \hline
\rowcolor{lightgray} {\sffamily\bfseries universal theories} & {\sffamily\bfseries interactions} \\ [1mm] \hline
$\bullet$ quantum theory & $\bullet$ gravity \\
$\bullet$ Special Relativity & $\bullet$ electromagnetism \\ 
$\bullet$ General Relativity & $\bullet$ weak interaction \\
$\bullet$ many particle physics & $\bullet$ strong interaction \\ [1mm]\hline
\rowcolor{lightgray} {\bfseries Problem:} Incompatibility between General Relativity and quantum theory & {\bfseries Wish:} Unification of all interactions  \\ [1mm]\hline
\multicolumn{2}{|>{\columncolor{black}}c|}{\color{white} {\bfseries Possible solution:} Quantum Gravity} \\ [1mm]\hline
\end{tabular} \label{Theories}}
\end{table}

A big problem in the theoretical description of physics is the incompatibility of quantum theory and GR as relativistic theory of gravity. This can be seen from the fact that in GR from very general assumption singularities, spatial points where all the surrounding matter fall into, will occur. Such a localization of matter is forbidden by quantum theory. Such incompatibilities make it necessary to look for a unification of quantum theory and GR, that is, for a quantum gravity theory (see, e.g.  \cite{GiuliniKieferLaemmerzahl03}, for a recent review). There are several approaches to such a new theory, e.g., string theory, canonical quantum gravity, or non--commutative geometry. In any case, deviations from standard physics given by the theories of Table \ref{Theories} are predicted. Each of these theories predicts deviations, e.g., from the Universality of Free Fall, modifications of the electromagnetic and of the gravitational interaction in terms of, e.g., a Yukawa--like potential, etc. All these theoretical considerations are strong reasons to make an effort to get better and better experimental results. Apart from that, physics always requires the best experimental foundation of its basic theories being the pillars of modern understanding of physics. 

As already mentioned, gravity plays an outstanding role. It plays this role not only because of the reasons described above but also because in most cases a violation of the principles underlying GR can be observed if the description of interactions like the Maxwell equationsm or the Dirac equation underlying particle physics are modified. Indeed, only the present form of Maxwell equations and Dirac equation is compatible with GR, that is, the structure of the Maxwell and Dirac equation strongly determines the structure of the gravitational field (a modification of Maxwell's equation, for example, leads to a violation of the Universality of Free Fall\cite{Ni77}). Consequently, tests of GR play an outstanding role in this search for new physics. 

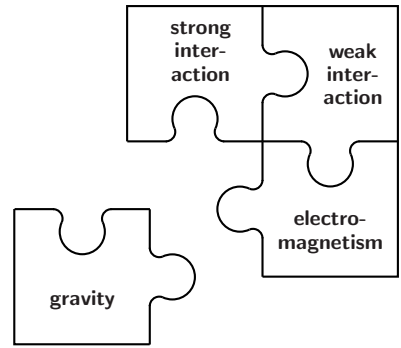
\begin{wrapfigure}[15]{R}{5.3cm}
\begin{minipage}[t]{5.4cm}
\psset{unit=0.15cm}
\begin{pspicture}(-12,-10)(22,22)
\rput{90}(10,12){
\rput(-6,0){\psarc(0,2){2}{-25}{205}
\psarc(-2.59,0.84){0.84}{-90}{22}
\psarc(2.59,0.84){0.84}{158}{270}
\psline(2.5,0)(6,0)
\psline(-2.5,0)(-6,0)}
\rput{180}(6,0){\psarc(0,2){2}{-25}{205}
\psarc(-2.59,0.84){0.84}{-90}{22}
\psarc(2.59,0.84){0.84}{158}{270}
\psline(2.5,0)(6,0)
\psline(-2.5,0)(-6,0)}
\rput{-90}(0,6){\psarc(0,2){2}{-25}{205}
\psarc(-2.59,0.84){0.84}{-90}{22}
\psarc(2.59,0.84){0.84}{158}{270}
\psline(2.5,0)(6,0)
\psline(-2.5,0)(-6,0)}
\rput{90}(0,-6){\psarc(0,2){2}{-25}{205}
\psarc(-2.59,0.84){0.84}{-90}{22}
\psarc(2.59,0.84){0.84}{158}{270}
\psline(2.5,0)(6,0)
\psline(-2.5,0)(-6,0)}
\psline(-12,0)(-12,-12)(12,-12)(12,12)(0,12)
\rput{180}(-12,10){\psarc(0,2){2}{-25}{205}
\psarc(-2.59,0.84){0.84}{-90}{22}
\psarc(2.59,0.84){0.84}{158}{270}
\psline(2.5,0)(6,0)
\psline(-2.5,0)(-6,0)}
\rput{90}(-6,16){\psarc(0,2){2}{-25}{205}
\psarc(-2.59,0.84){0.84}{-90}{22}
\psarc(2.59,0.84){0.84}{158}{270}
\psline(2.5,0)(6,0)
\psline(-2.5,0)(-6,0)}
\psline(-18,10)(-18,22)(-6,22)
\rput{-90}(-8,-6){\txt\footnotesize\sffamily\bfseries{electro- \\ magnetism}}
\rput{-90}(6,-8){\txt\footnotesize\sffamily\bfseries{weak \\ inter- \\ action }}
\rput{-90}(8,5.5){\txt\footnotesize\sffamily\bfseries{strong \\ inter- \\ action }}
\rput{-90}(-14,16){\txt\footnotesize\sffamily\bfseries{gravity}}
}
\end{pspicture}
\end{minipage}
\caption{Until now, gravity does not fit into the unification scheme.}
\end{wrapfigure}

Behind all these physical structures there are the principles of many particle physics. Today, this field is deeply connected with renormalization group theory. Due to the fact that renormalization group theory not only is a method describing the physics of gases and their phase transitions but, more generally, also is a mathematical method with applications in many parts of physics, from statistical physics to hydrodynamics, solid state physics to elementary particle physics, and even to problems in quantum gravity, it is very important to understand the principles underlying this theory more deeply and to improve the quality of its tests. 

\subsection{Structure of gravity}

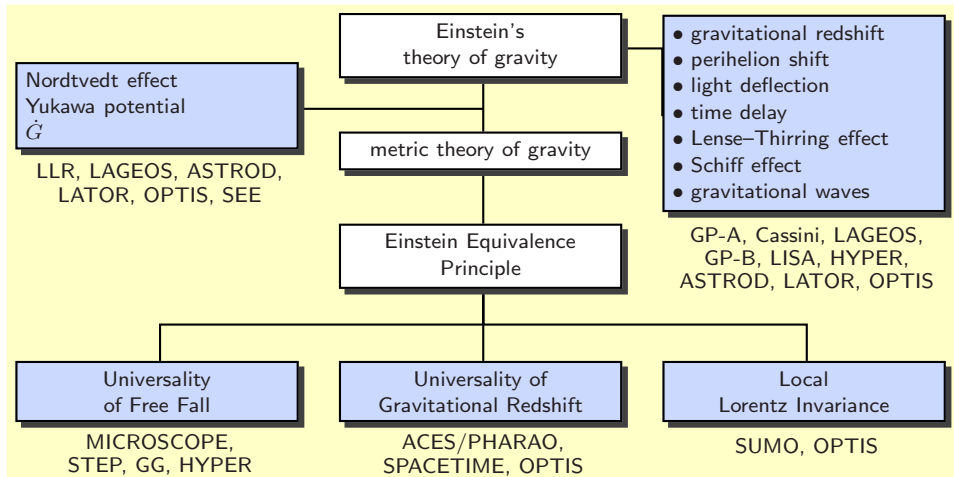
\begin{figure}[t]
\psset{xunit=1cm}
\begin{center}
{\footnotesize\sffamily
\begin{pspicture}(-6,-1)(6,5)
\psframe[linestyle=none,fillstyle=solid,fillcolor=light-yellow](-0.5\textwidth,-1.4)(0.5\textwidth,5)
\rput(0,4.4){\rnode{ET}{\ShowX{white}{3.5cm}{Einstein's \\ theory of gravity}}}
\rput(0,3){\Rnode{MT}{\ShowX{white}{3.5cm}{metric theory of gravity}}}
\rput(0,1.6){\Rnode{EEP}{\ShowX{white}{3.5cm}{Einstein Equivalence Principle}}}
\rput(-4.3,-0.2){\Rnode{UFF}{\ShowX{light-blue}{3.5cm}{Universality \\ of Free Fall}}}
\rput(0,-0.2){\Rnode{UGR}{\ShowX{light-blue}{3.5cm}{Universality of \\ Gravitational Redshift}}}
\rput(4.3,-0.2){\Rnode{LLI}{\ShowX{light-blue}{3.5cm}{Local \\ Lorentz Invariance}}}
\rput(4.3,3.5){\rnode{LT}{\ShowL{light-blue}{3.5cm}{$\bullet$ gravitational redshift \\ $\bullet$ perihelion shift \\ $\bullet$ light deflection \\ $\bullet$ time delay \\ $\bullet$ Lense--Thirring effect \\ $\bullet$ Schiff effect \\ $\bullet$ gravitational waves}}}
\psline(-4.3,3.6)(0,3.6)
\rput(-4.3,3.6){\ShowL{light-blue}{3.5cm}{Nordtvedt effect \\ Yukawa potential \\ $\dot G$}}
\rput(-4.3,-1){\txt{MICROSCOPE, \\ STEP, GG, HYPER}}
\rput(0,-1){\txt{ACES/PHARAO, \\ SPACETIME, OPTIS}}
\rput(4.3,-0.9){SUMO, OPTIS}
\rput(4.3,1.6){\txt{GP-A, Cassini, LAGEOS, \\ GP-B, LISA, HYPER, \\ ASTROD, LATOR, OPTIS}}
\rput(-4.3,2.6){\txt{LLR, LAGEOS, ASTROD, \\ LATOR, OPTIS, SEE}}
\ncangle[angleA=90,angleB=-90,arm=0.4cm]{MT}{ET}
\ncangle[angleA=90,angleB=-90,arm=0.4cm]{EEP}{MT}
\ncangle[angleA=90,angleB=-90,arm=0.4cm]{UFF}{EEP}
\ncangle[angleA=90,angleB=-90,arm=0.4cm]{UGR}{EEP}
\ncangle[angleA=90,angleB=-90,arm=0.4cm]{LLI}{EEP}
\ncangle[angleA=0,angleB=180,arm=0cm]{ET}{LT}
%
\end{pspicture}}
\end{center}
\medskip
\caption{The structure of experimental exploration of the theory of gravity. Testable issues are described in grey boxes, theoretical concepts in white boxes. Metric theories of gravity are based on the directly testable principles Universality of Free Fall, Universality of the Gravitational Redshift, and Local Lorentz Invariance. Particular effects, like the Nordtvedt--effect, a time--varying $G$, a deviation from ordinary Newton potential denote deviations from ordinary Einsteinian General Relativity. The predictions of Einstein's GR are found in the upper right box. The missions aiming at the exploration of the various effects are shown below the grey boxes.  \label{Scheme}}
\end{figure}

The present theory of gravity, Einstein's General Relativity, is based on a set of universality principles\cite{Will93}, (i) the Universality of Free Fall (UFF), (ii) the Universality of the Gravitational Redshift (UGR), and (iii) a universality with respect to the state of motion of the observer, called Local Lorentz Invariance (LLI). If these principles are valid, then gravity can be described by a space--time metric as given in the mathematical framework of Riemannian geometry. Further requirements on the structure of the equations the metric has to fulfill then lead to the Einstein field equation, see Fig.\ref{Scheme}. 

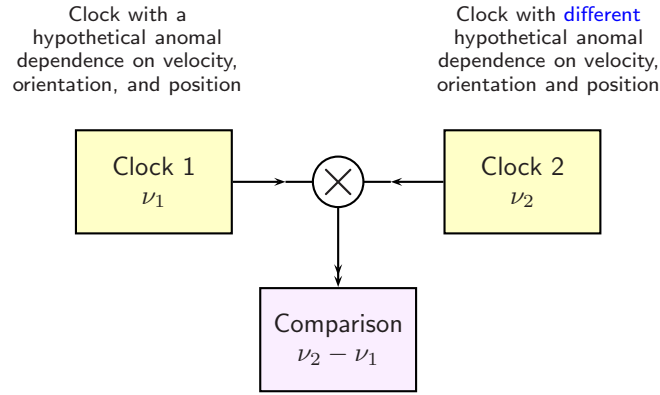
\begin{figure}[t]
\begin{center}
\psset{unit=0.7cm}
\begin{pspicture}(-6,-3)(6,5)
\psframe[fillstyle=solid,fillcolor=light-yellow](-5,0)(-2,2)
\rput(-3.5,1){\txt\sffamily{Clock 1 \\ $\nu_1$}}
\rput(-4,3.5){\txt\footnotesize\sffamily{Clock with a \\ hypothetical anomal \\ 
dependence on velocity, \\ orientation, and position }}
\psframe[fillstyle=solid,fillcolor=light-yellow](2,0)(5,2)
\rput(3.5,1){\txt\sffamily{Clock 2 \\ $\nu_2$}}
\rput(4,3.5){\txt\footnotesize\sffamily{Clock with \textcolor{blue}{different} \\ 
hypothetical anomal \\ dependence on 
velocity, \\ orientation and position}}
\psline{->}(-2,1)(-1,1)
\psline{->}(2,1)(1,1)
\psline{->>}(-1,1)(0,1)(0,-1)
\psline(1,1)(0,1)
\pscircle[fillstyle=solid,fillcolor=white](0,1){0.5}
\rput(0,1){\huge $\times$}
\psframe[fillstyle=solid,fillcolor=llight-red](-1.5,-3)(1.5,-1)
\rput(0,-2){\txt\sffamily{Comparison \\ $\nu_2 - \nu_1$}}
\end{pspicture}
\end{center}
\caption{General scheme for testing SR and GR with clocks. Tests of the constancy of the speed of light and of UGR are essentially carried through with clocks whose frequency may change with orientation, velocity, and position. If furthermore, the clocks are assumed to move with different velocities or to be at different positions, then this scheme also applies to tests of the time dilation and of the gravitational redshift. \label{Fig:clocks}}
\end{figure}

Indeed, most fundamental physics experiments are devoted to tests underlying the principles of GR.  Owing to the fact that GR deals with the structure of space--time, all tests of GR are tests involving the measurement of time, of paths (either paths of light or paths of massive bodies), and of directions. Here, clocks play a really fundamental role: they are used for a complete test of the principles underlying SR and also for tests concerning the gravitational redshift, see Fig.\ref{Fig:clocks}. That is also the reason why each improvement in the precision of clocks is followed by new tests of SR and GR. The observation of paths, that is, geodesic paths driven by the gravitational interaction only, is difficult due to many disturbing and competetive effects. That effects are (i) atmospheric drag and radiation from the Earth and the Sun which lead to non--gravitational accelerations, and (ii) effects due, e.g., to gravitational multipole moments of the Earth and the Sun which are not known to the needed precision or which need a complicated data analysis. Only GP-B tests the parallel transport of a direction given by a distinguished physical system, namely gyroscopes. 

\section{Fundamental Quests}

As already mentioned, all approaches to a quantum gravity theory predict deviations from the principles underlying GR asking for refined tests of all the aspects of GR. More specific, these experiments look for 
\begin{itemize}
\item violations of the UFF
\item violations of the UGR
\begin{itemize}
\item time--dependence of the fine structure constant
\item time--dependence of the gravitational constant
\end{itemize}
\item violations of LLI in many aspects: 
\begin{itemize}
\item non--isotropy of light propagation
\item non--constancy of velocity of light
\item fundamental dispersion of light propagation
\item non--isotropy of elementary particle parameters like mass
\item search for anomalous spin--interactions
\end{itemize}
\item non--Einsteinian effects like
\begin{itemize}
\item Yukawa--like gravitational potential
\item Nordtvedt--effect
\item time--variation of the gravitational constant $G$
\end{itemize}
\end{itemize}
The corresponding space missions can be found in Fig.\ref{Scheme}. Beside tests of relativity and gravity, there are also issues for testing the fundamentals of the other universal theories like quantum theory and many particle theory:
\begin{itemize}
\item linearity of quantum physics
\item entanglement
\item Casimir force
\item physics of Bose--Einstein condensates
\item search for a fundamental decoherence
\item test of renormalization group theory
\end{itemize}

Since quantum gravity is characterized by the Planck energy $E_{\rm P} \sim 10^{28}\;\hbox{eV}$ and laboratory energies are of the order eV for ordinary, e.g., optical laboratory experiments to GeV for large particle accelerators, quantum gravity effects are too small by many orders of magnitude to be detectable in laboratories. Indeed, expected quantum gravity induced violations of LLI, for example, are looked for in ultra heigh energy cosmic rays. However, all the present predictions from quantum gravity are in fact merely hypotheses, they are not based on complete theories. Furthermore, perhaps some additional mechanism has to be applied which may lead to some enhancement of the expected effect as it is the case for deviations from Newton potential at small distances as predicted from higher dimensional theories. Therefore, there is always a possibility that deviations from standard physics may occur at lower energies than given by the Planck scale. Consequently, {\it any improvement of the accuracy of experimental results is of great value}.

We also address the question what happens if an experiment shows a violation of one of these basic principles. This not necessarily means that one has found a violation of one of these principles. This effect may also be a result of a new interaction which might well be in accordance with the tested principles. Therefore one first has to search whether this effect can be shielded, or whether one can find a cause of this effect in the sense of a source which creates a field causing this effect. In both cases the effect can be considered to be a new interaction. Only if all these questions are answered appropriately one can speak about a violation of a basic principle. 

Particular predictions of deviations from standard physics are
\begin{itemize}
\item Violation of the UFF at the $10^{-13}$ level predicted from dilaton scenarios\cite{DamourPiazzaVeneziano02}.
\item Deviation from GR in terms of the PPN parameters $\gamma$ and $\beta$, again predicted within dilaton scenarios\cite{DamourPiazzaVeneziano02a}. 
\item Violations of LLI at Planck scale predicted from non--commutative geometry aproaches\cite{Carrolletal01}
\item Additional Yukawa part of the gravitational potential at small distances predicted from higher--dimensional theories\cite{Antoniadis03}.
\end{itemize}

\section{The space conditions}

In many cases the sensitivity of measuring devices and/or the accuracy of measurements itself will increase if the experiments can be performed under
conditions of free fall, that is, under conditions of weightlessness. The
advantages of such conditions are:

\begin{enumerate}
\item  The infinitely long, and periodic, free-fall. As an example, long free
fall conditions enable high precision tests of the UFF
for all kinds of structureless (i.e. pointlike) matter.

\item  Long interaction times: This is, for example, a big advantage in
atomic or molecular interferometers, where the laser cooled atoms or molecules may interact
with other external fields for a long time and do not fall out of the interferometer as it happens on Earth. Only in a microgravity environment in space one has the opportunity of a dedicated long exposure to certain interactions. 

\item  Large potential differences. In a large class of experiments (e.g. tests
of the gravitational redshift), the magnitude of the signals looked for depends on the
difference of the positions of the clocks in the gravitational potential. It is obvious that this can be achieved best by going into space.

\item  Large velocity changes. For macroscopic devices used for, e.g., testing the dependence of the speed of light with respect to the laboratory velocity (Kennedy-Thorndike-tests), the
maximum velocity on Earth might be of the order 1000 km/h. In space this
can be increased by about one order of magnitude. For example, the velocity
variations along the orbit (e.g. in a high elliptical Earth orbit) are 30
times higher than one can attain using the Earth's rotation.

\item  Availability of long distances. In space, much longer distances are
available than in any laboratory on Earth. This is essential, e.g.,
in the study of low frequency (10$^{-3}$ Hz) gravity waves using
interferometric techniques, where the strain of spacetime is to be measured at
or below the 10$^{-21}$ level.

\item  A low noise / vibration environment. Seismic noise is a limiting factor
for many experiments on Earth (e.g. for gravitational wave detectors and for
torsion balances) in the frequency range below 10 Hz.

\item For certain interactions, only in space one has the opportunity to search for the corresponding effects. As an example, only in space there are appropriate conditions adapted to the detection of the gravitational time delay, of gravitational waves with very low frequencies, of the Schiff-- or of the Lense--Thirring effect. 

\item Due to the absence of the atmosphere, the true particle content in outer space is directly observable. 
\end{enumerate}

As a consequence, there are quite a few instances where it is really indispensible to go to space in order to achieve improved accuracies as compared to experiments on ground. 

However, these advantages have to be compared with some disadvantages. These disadvantages are in many cases 
\begin{enumerate}
\item huge financial effort,
\item long time for preparation,
\item no direct access to the experiment during operation,
\item no possibility of a post--mission analysis of the experimental payload.
\end{enumerate}
In the case of experiments to be performed onboard of the ISS, the last two disadvantages are not present. At least after the mission an extensive post--mission check can be carried through. 

As far as experiments on the ISS are concerned, the situation in general is a bit different from that for dedicated satellite missions. It is clear that there are some disadvantages due to the very existence and construction of the ISS: Due to the atmospheric drag, the true free-fall
inside the ISS is rather short and, due to the circular orbit, the difference in
the gravitational potential of the Earth is small. In addition, the large
structure and movable parts on the ISS create a rather large vibrational
noise, and the non--negligible Earth's gravitational gradient as well as the
gravitational field of the ISS itself give a rather high level of
residual acceleration. Therefore, many high--precision experiments cannot be performed on the ISS. However, the ISS may be used as an important and very
appropriate test bed for certain dedicated Fundamental Physics mission
satellites. 

On the other hand, the ISS environment enables experiments to be
conducted in a way that would be quite impossible using satellites. Due to the
regular servicing of the space station, exchange, repair, and improvements of
experimental facilities on board the ISS are possible. Facilities also can be
brought back to Earth for post-mission analysis of effects that may have been
causing potential systematic errors, and, from a physical point of
view, it is of prime importance to have the capacity to repeat experiments,
and to test the reproducibility of results. Undeniably, one of the most
powerful arguments for the utilization of the ISS for fundamental physics experiments,
notwithstanding the less than ideal environmental conditions on board, is the
unrivalled opportunity for quicker and easier access to the experimental
apparatus than is conceivably possible using dedicated satellites. 

\section{Past and running missions}

Until now, there have been only very few space missions which carried through dedicated experiments concerning gravitational physics or which could be used for that. Nevertheless, the results of these missions have been widely discussed and led to a much better experimental basis and understanding of gravitational physics. Without these missions some effects like the gravitational redshift or the gravitational time delay would have been be confirmed with much lower accuracy, and some effects even would still remained unobserved like the geodetic precession of the Earth--Moon system or the Lense--Thirring effect. Already from this one can infer the scientific potential of further space missions. 

\noindent {\bf GP-A.} This was the first space mission dedicated to fundamental physics issues. The time given by a H--maser on a rocket was compared with the time of a ground based H--maser. Due to a three--channel method, one could separate between the Doppler effect from the gravitational redshift. This yielded the up to now best confirmation of the absolute gravitational redshift with an accuracy of $1.4 \cdot 10^{-4}$.

\smallskip
\noindent {\bf Viking.} Due to a microwave link to the Mars explorer one could measure the time needed for a signal to propagate from Earth to the Viking satellite and back. In case of a conjunction, the travel time should be longer according to GR. This could be seen with an accuracy of almost $10^{-3}$. Competing and limiting effects came from the Solar and Earth's atmosphere. 

\smallskip
\noindent {\bf LLR.} Various Apollo Moon missions as well as Russian unmanned Moon missions placed several laser retroreflectors on the moon. Laser tracking from the Earth yielded after 20 years for the determination of the Earth--Moon distance an accuracy of about 1 cm. Using these data, effects like the geodetic precession of the Earth--Moon system could be verified, the validity of the strong Equivalence Principle (UFF together with the gravitational self energy) could be tested, and a comparison of the data with a hypothetical Newtonian Yukawa potential yielded strong estimates on such effects. 

\smallskip
\noindent {\bf LAGEOS.} This mission consists of two passive small satellites orbiting the Earth which are laser tracked; it gives informations about the gravitational field of the Earth. The orbital data of these satellites have been used to experimentally check for the first time 
the existence of the gravitomagnetic Lense--Thirring effect with with a claimed 20 -- 30\% accuracy\cite{Ciufolinietal98}. The difficulty lies in competing effects from higher gravitational multipole which are actually much larger than the effect looked for. The basic idea of the data analysis is to combine the data of two satellites (LAGEOS and LAGEOS II) in such a way that the lowest order gravitational multipoles cancel. Any improvement of Earth gravity models\cite{IorioMorea04} as well as additional data from further satellites\cite{Iorioetal03} will improve the results (see the two talks of L. Iorio). 

\smallskip
\noindent {\bf Cassini.} The Cassini mission to Saturn is equipped with a multi--frequency radio--link to the Earth. Using this technique, the disturbing effects due to the Sun's corona for measurement of the travel time of signals from the Earth to Cassini and back could be removed almost completely. The achieved accuracy of $2 \cdot 10^{-5}$ is almost two orders of magnitude better than their previous Viking result. 

\section{Possible future missions}

Here we shortly describe the scientific objectives and other main issues of planned space missions devoted to fundamental physics. For most of these missions, more informations can be found in the reviews\cite{LaemmerzahlDittus02,Laemmerzahletal04} and the references cited therein. 

\smallskip
\noindent {\bf GP-B.} With this mission\cite{Everittetal01}, the gravitomagnetic Schiff or frame--dragging effect, shall be observed with 0.1\% accuracy\cite{Everittetal01}. The main part of the satellite is a huge dewar which maintains a cryogenic environment for 18 months. The main part of the experimental payload located inside the dewar are gyroscopes made from superconducting rotating spheres and a telescope. The gyroscopes represent the inertial systems which, due to the gravitomagnetic field of the rotating Earth start to precess. The precession is read out with SQUIDs using a magnetic field attached to the rotating spheres. This direction then is compared with the direction given by distant stars which is obsered by the telescope. This mission should start in 2004. See the talks by M. Keiser, R. Torii and Y. Ohshima. 

\smallskip
\noindent {\bf ACES/PHARAO.} In this mission\cite{Salomonetal01} the PHARAO clock, based on a fountain of cold cesium atoms, and a hydrogen maser clock will be brought onboard the ISS, complemented by a microwave link for synchronization with clocks on Earth. With this equipment, better tests of the gravitational redshift can be performed and one can search for a time--dependence of the fine structure constant at the $10^{-16}$ level. Furthermore, by establishing such clocks in space represents an enormous improvement over the present level of synchronization that is possible using GPS clocks. 

\smallskip
\noindent {\bf PARCS.} This ISS project (see, e.g., the review \cite{Laemmerzahletal04}) consists, similar to PHARAO, of a cesium atomic clock which, together with SUMO, will test the Universality of the Gravitational Redshift, the constancy and isotropy of the speed of light and intends to establish a better time standard by a factor of 20. The accuracy of PARCS will be $10^{-16}$. 

\smallskip
\noindent {\bf RACE.} For RACE, a double MOT (Magneto Optical Trap) design is used to multiply launch Rb atoms, which, as compared to Cs, possess a much lower collision shift error. Furthermore, RACE uses two cavities in order to interrogate the atomic frequency. Among others, one advantage of that design is the possibility to eliminate the vibrational noise of the ISS. RACE will aim at a clock accuracy of $10^{-17}$. 

\smallskip
\noindent {\bf MICROSCOPE.} This already approved mission\cite{Touboul01a} is devoted to a test of the UFF to an accuracy of $10^{-15}$ in terms of the E\"otv\"os parameter. The relative accelerations of two pairs of test masses are measured: the first pair consists of a Pt/Rh alloy and special Ti alloy (TA6V), and the second pair consists of two identical Pt/Rh alloys. The second pair is taken for redundancy and control. The relative acceleration is measured using capacitive sensors. 

\smallskip
\noindent {\bf LISA.} With a huge two--arm interferometer made up of three satellites being 5 million km apart one will detect gravitational waves in the frequency range of $10^{-1}$ to $10^{-4}$ Hz. Each of these satellites carries two phase--locked laser systems and two mirrors. Each mirror is controlled by a drag--free system in order that the satellite very precisely moves on a geodesic path within the timescale of 1000 s. LISA has the status of an ESA--NASA cornerstone mission. 

\smallskip
\noindent {\bf SUMO.} Three superconducting microwave oscillators, placed in the Low Temperature Microgravity Physics Facility (LTMPF) on the ISS, are designed to perform better tests of the isotropy of the speed of light. Furthermore, a fiber--optic link with PARCS, an atomic clock on the ISS, will enable tests of the constancy of the speed of light and the UGR for the comparison resonator -- atomic clock. Integration over a few months will give improvements on the isotropy of the velocity of light by two orders of magnitude, and for the constancy of the speed of light by three orders of magnitude\cite{Laemmerzahletal04}. 

\smallskip
\noindent {\bf SUE, BEST, MISTE, DYNAMX, EXACT.} These projects deal with the physics of liquid Helium and are planned to be carried through on the ISS. The planned tests are dedicated to the understanding of the formation of long--range order below a sharply defined temperature. The tests cover various universality aspects of renormalization group theory of many particle systems near phase transitions. (Due to lack of space--time we do not treat them separately, more information can be found in a recent review\cite{Laemmerzahletal04}.)

\smallskip
\noindent {\bf STEP.} This mission\cite{Lockerbieetal01} wants to test the UFF to a precision of $10^{-18}$. In contrast to MICROSCOPE it uses cryogenic techniques: SQUIDs are used to determine the relative position between the freely flying test masses. Four pairs of test--masses are used, that is, the pair Pt/Ir -- Nb, the pair Nb -- Be, and two pairs Pt/Ir -- Be, so that there is a  redundancy as well as a cyclic condition for which the total acceleration difference between pairs of test masses must add to zero in the case that UFF holds. The high accuracy requires several additional techniques. First, the test masses must have an appropriate design in order to be insensitive to gravity gradients, (see also the talk of R. Torii and some remarks of caution by B. Lange). 

\smallskip
\noindent {\bf ASTROD.} This is a proposed Chinese interplanetary laser ranging mission\cite{Nietal02} which aims at (i) an improvement of the determination of the PPN parameters $\gamma$ and $\beta$ by three to six orders of magnitude, (ii) the detection of gravitational waves below the mHz range, and (iii) to improve the knowledge of solar system parameters like the angular momentum of the Sun and asteroid masses. The main technique is laser ranging for which new techniques for the coupling of very weak laser light to local oscillators has to be developed. 

\smallskip
\noindent {\bf LATOR.} This recently proposed\cite{TuryshevShaoNordtvedt03,TuryshevShaoNordtvedt04} interplanetary ranging mission aimes at measuring the deflection of light with a precision of $10^{-8}$. The main idea is to have two small spacecrafts and a reference on the ISS spanning a triangle and to measure the lengths of the three sides of this triangle and, in addition, the observed angle of the light rays from the satellites to the ISS. From the gravity induced deformations of ordinary Euclidean geometry of the  paths of light one can infer the gravitational influence. The aimed high precision can be achieved by an optical truss provided by a 100 m long multi--channel optical interferometer mounted on the ISS. Beside the deflection of light, also second order effects, the Sun's quadrupole momentum and the Lense--Thirring effect can be measured. 

\smallskip
\noindent {\bf OPTIS.} This mission aims at an improvement of quite a few tests of SR and GR, namely (i) the isotropy of the speed of light, (ii) the constancy of the speed of light, (iii) the Doppler effect (or time dilation), (iv) tests of the UGR comparing various atomic clocks, (v) tests of UGR comparing an optical resonator and atomic clocks, (vi) measurement of the absolute gravitational redshift, (vii) measurement of the Lense--Thirring effect, (viii) measurement of the perigee advance, and (ix) test of Newton's $1/r$ potential. The main components of the experimental payload are clocks (resonators and atomic clocks (H--maser, ion clocks)), a laser link to the Earth, and a drag-free control of the satellite. More information can be found in a recent preprint\cite{Laemmerzahletal04a}, (see also the talk of S. Schiller).

\smallskip
\noindent {\bf SPACETIME.} With this mission\cite{MalekiPrestage01} three ion clocks with stability of the order $10^{-16}$ will be brought to the Sun as close as about 5 Solar radii. From the comparison of the these highly accurate clocks during their motion through a strongly changing gravitational potential one gets a huge improvement of tests of the UGR. The errors can be reduced considerably by the possibility to place all three clocks in the same environment.

\smallskip
\noindent {\bf GG.} Here again the UFF is aimed to be tested at an accuracy of $10^{-17}$. The main idea of this Italian proposal\cite{WEbpage} is that a high frequency modulation of the UFF--violating signal induced by a rotation of the test mass can improve the signal--to--noise ratio. 

\smallskip
\noindent {\bf SEE.} In this mission\cite{Sandersetal00}, two freely falling interacting test masses are placed in a big container orbiting the Earth in free fall. Using a newly invented and highly precise device for monitoring the positions of the two masses, one can test the (i) validity of the Newton $1/r$ potential over distances between the two masses and between the Earth and the satellite. Furthermore, one can make (ii) better tests of the UFF, (iii) more precise measurements of the gravitational constant, and (iv) also search for a time--dependence of the gravitational constant. 

\smallskip
\noindent {\bf HYPER.} This missions aims at (i) measuring the Schiff effect, (ii) testing the UFF, (iii) search for a fundamental decoherence in quantum mechanics, (iv) more precise measurement of the fine structure constant. The main feature of this mission is that it wants to employ atomic interferometry in space. A more technical aspect is that with this mission accelerometers (gravity reference sensors) and gyroscopes based on atomic interferometry might be introduced. Working with atomic interferometry, prerequisite techniques are lasers, laser cooling, and magneto--optical traps (see the contribution of C. Jentsch). 

\section{Key techniques}

\subsection{Key space techniques}

\noindent {\bf Drag--free control.} This technique is necessary to assure that the satellite moving as close as possibile along a geodesic path given by gravity only. It needs the interplay between very sensitive inertial sensors and very precise microthrusters. For this purpose, algorithms have to be developed which process the signals from the sensors and precisely drive the microthrusters, where their corresponding noise properties have to be taken into account. The corresponding control system is called Drag--Free and Attitude Control System (DFACS).

\smallskip
\noindent {\bf Gravity reference sensors (Inertial sensors).} In the last years so-called 'drag-free sensors' have been developed that offer the opportunity to
cancel out all non-gravitational disturbing forces and torques (like
air--drag, solar pressure, magnetic field disturbances etc) on 
orbiting satellites. The drag--free concept involves centering a free-floating test
mass located inside a satellite which is free of external disturbances and follows a purely gravitational orbit. External (non--gravitational) forces and
torques will move the satellite relative to the test mass. The change
in the relative position is measured and then used to derive the
appropriate control force which has to be applied by the DFACS in
order to drive the test mass displacement to zero. Since the satellite is forced to follow the test mass, it follows the same gravitational orbit.

The test mass displacement can be measured electrostatically or magnetically. In most cases (e.g., for MICROSCOPE), a capacitive method is used where the test mass is surrounded by electrodes. One area of the test mass and one electrode form a capacitor and the displacement induced change of its capacity can be measured, see e.g. \cite{Touboul01}. 

\smallskip
\noindent {\bf Microthrusters.} Microthrusters are needed for very precise navigation of satellites. Field Electric Emission Propulsion (FEEP) ion thrusters or colloidal thrusters are used to control the residual acceleration down to $10^{-10}\;{\hbox{m/s}}^2$ in the signal bandwidth. This sets an upper limit for the thrust: linear forces acting on the satellite are less than $50\; \mu \hbox{N}$ in all 3 axes and maximum torques are about $10\; \mu \hbox{Nm}$. The resolution of thrust control has to be done with an accuracy of about $0.1\; \mu \hbox{N}$. For a satellite diameter of about 1.5 m the solar radiation pressure of about $4.4 \;\mu \hbox{N}/\hbox{m}^2$ and the radiation pressure of the Earth albedo of $1.2\; \mu \hbox{N}/\hbox{m}^2$ sum up to a total drag of about $10 \;\mu \hbox{N}$. 


\smallskip
\noindent {\bf Lasertracking.} In order to measure the distance between a ground station and the satellite (or another object) with very high precision the Satellite Laser Ranging (SLR)
technique can be used. This is important for LLR and LAGEOS and will be used in OPTIS, LATOR, ASTROD. Very short laser pulses are transmitted
from a telescope in a ground station to a satellite, from which they are retro-reflected back to the station by a corner cube reflector. The round trip time is measured and gives the distance.
In other words, SLR measures the absolute time of flight of photons so that the geometry of satellite and laser station can be determined
precisely as long as the system calibration error is controlled to a
negligible level. The present state of the art is that for the travel time accuracies of up to 50 picoseconds or better can be measured. This is equivalent to an accuracy of 1 centimeter or less.
Currently, NASA is building up the Satellite Laser Ranging 2000 system.
SLR2000 is an autonomous and eyesafe photon--counting SLR station with
an expected single shot range precision of about one centimeter and a
normal point precision better than 3 mm. The system will provide
continuous 24 hour tracking coverage of artificial
satellites at altitudes up to 20 000 km. Approximately forty laser station systems, distributed all over the world, now contribute to this technique. These stations form a network that is coordinated by the International Laser Ranging  Service: ILRS and by a European consortium EUROLAS.

\smallskip
\noindent{\bf Star trackers.} Star Tracker are sensors that are used in satellite attitude control in order to achieve accurate pointing measurements. The sky is scanned by e.g. a CCD camera. From the detected star patterns a computer algorithm can determine the pointing direction of the sensor and thus of the spacecraft. Today state-of the-art high precision star trackers can 
provide a single star angular accuracy better than 3 arc seconds.

\subsection{Key payload techniques} Here we describe a few key payload techniques which already played a role in missions or which are planned to be part of future missions. 

%

\smallskip
\noindent {\bf Clocks.} All kinds of clocks will play a major role in future fundamental physics missions. The reason for that is that most of these tests are clock--comparison tests: Tests of LLI consist in comparison of clocks with different internal orientations, of clocks with a different hypothetical velocity dependence (Kennedy-Thorndike tests) and clocks in different states of motion. All redshift experiments compare clocks at different positions in the gravitational field. We shortly list the types of clocks:

\smallskip
\noindent{\it H--maser.} H--masers are based on a hyperfine transition of the ground state of the hydrogen atom with a life time of about 1 s which is coupled to a resonator. The frequency is 1.420 405 751 Hz and the Allan deviation is less than $10^{-15}$. H--masers are already space qualified. 

\smallskip
\noindent{\it Atomic fountain clocks}
 \smallskip
\noindent{\it Ion clocks.} Today, ion clocks\cite{Prestageetal01} approach the level of $10^{-16}$ in their stability (in terms of the Allan variance). Ion clocks are based on hyperfine transitions of trapped ions (e.g. ${\hbox{Hg}}^+$, ${\hbox{Cd}}^+$, ${\hbox{Yb}}^+$). 

\smallskip
\noindent{\it Resonators.} Resonators (cavities) are a realization of so--called light clocks. Locking of lasers to optical resonators (covered by the talk of S. Schiller) used for the best Michelson--Morley tests\cite{Muelleretal03c} will give highly stabilized frequencies which carry, via $\nu = n c/(2L)$, where $L$ is the length of the cavity, the information about the velocity of light $c$ for propagation along the cavity axis. This information is used in order to make statements about the isotropy and the constancy of the speed of light. For this, the length of the cavity has to be very stable since otherwise this could mask the searched effect. For cryogenic resonators the stability is $\delta L/L \leq 7 \cdot 10^{-16}$ \cite{Muelleretal03c} what corresponds to the $30^{\rm th}$ part of the diameter of the proton. The high dimensional stability requires materials with low thermal expansion coefficients. Another requirement is stable lock of lasers to the cavity which can be obtained. --- Microwave resonators work in the same way, the only difference is the longer wavelength (in the cm range) of the electromagnetic radiation in the resonator. A particular development in this area are the whispering gallery resonators where the resonator has cylindric geometry and the radiation possesses large angular mode numbers. These have been used for the today's best test\cite{Wolfetal04} of the constancy of $c$ (see the talk of M. Tobar).

\smallskip
\noindent Today the most advanced clocks approach an accuracy of $10^{-16}$. From the gravitational redshift, it can be seen that these clocks run differently if they are located at a height difference of 90 cm. If the precision of clocks will improve by one or two orders, then, due to the fact that the surface of the Earth is not really constant, these clocks cannot define a well defined timne. For that, one has to go to space. Only in space the conditions are well defined enough in oder that such high precision apparatus can yield unique and interpretable results. 

\smallskip
\noindent {\bf Lasers.} There are already space proven lasers available. These lasers are diode--pumped Nd:YAG lasers. They possess high intensity and frequency stability. Lasers will be applied in the missions LISA, HYPER, OPTIS, LATOR, and ASTROD, indicating the overall importance of this device in space technology.

\smallskip
\noindent {\bf Frequency combs.} Tests of the constancy of $c$ and of the UGR with optical resonators require a high--precision technique for comparing frequencies in the microwave and optical range, differing by more than 5 orders of magnitude. The recently invented  frequency comb is the appropriate technique, see \cite{CundiffYeHall01} for an overview. This technique is simpler, cheaper, power--saving and more accurate than previous methods. In the laboratory, the comparison of the frequencies with an accuracy of $10^{-15}$ has already been achieved.  

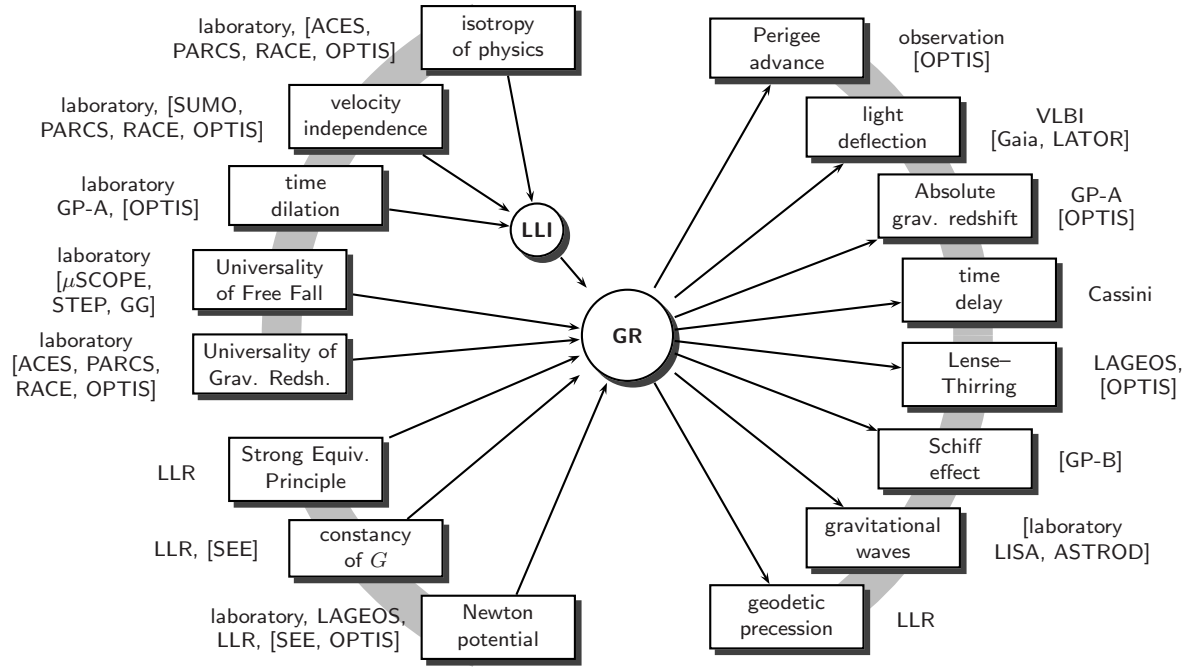
\begin{sidewaysfigure}
\psset{xunit=0.8cm,yunit=0.7cm}
\begin{center}
{\footnotesize\sffamily
\begin{pspicture}(-10,-7)(9,6)
\psarc[linewidth=15pt,linecolor=lightgray](0,0){4.6}{0}{60}
\psarc[linewidth=15pt,linecolor=lightgray](0,0){4.6}{300}{360}
\psarc[linewidth=15pt,linecolor=lightgray](0,0){4.6}{115}{190}
\psarc[linewidth=15pt,linecolor=lightgray](0,0){4.6}{200}{245}
\rput(0,0){\rnode{GR}{\pscirclebox[shadow=true]{\txt\bfseries{\;\; GR \;\;}}}}
\rput(-1.5,2){\rnode{LLI}{\pscirclebox[shadow=true]{\txt\bfseries{LLI}}}}
\rput(-2.1,5.6){\rnode{LLI1}{\ShowX{white}{1.8cm}{isotropy \\ of physics}}}
\rput(-5.7,5.6){\txt{laboratory, [ACES, \\ PARCS, RACE, OPTIS]}}
%
\rput(-4.3,4.1){\rnode{LLI2}{\ShowX{white}{1.8cm}{velocity \\ independence}}}
\rput(-7.9,4.1){\txt{laboratory, [SUMO, \\ PARCS, RACE, OPTIS]}}
\rput(-5.3,2.6){\rnode{LLI3}{\ShowX{white}{1.8cm}{time \\ dilation}}}
\rput(-8.3,2.6){\txt{laboratory \\ GP-A, [OPTIS]}}
\rput(-5.9,1){\rnode{UFF}{\ShowX{white}{1.8cm}{Universality \\ of Free Fall}}}
\rput(-8.7,1){\txt{laboratory \\ [$\mu$SCOPE, \\ STEP, GG]}}
\rput(-5.9,-0.6){\rnode{UGR}{\ShowX{white}{1.8cm}{Universality of \\ Grav. Redsh.}}}
\rput(-9,-0.6){\txt{laboratory \\ [ACES, PARCS, \\ RACE, OPTIS]}}
\rput(-5.3,-2.6){\rnode{SEP}{\ShowX{white}{1.8cm}{Strong Equiv. \\ Principle}}}
\rput(-7.5,-2.6){\txt{LLR}}
\rput(-4.3,-4.1){\rnode{Gdot}{\ShowX{white}{1.8cm}{constancy \\ of $G$}}}
\rput(-7,-4.1){\txt{LLR, [SEE]}}
\rput(-2.1,-5.6){\rnode{NP}{\ShowX{white}{1.8cm}{Newton \\ potential}}}
\rput(-5.3,-5.6){\txt{laboratory, LAGEOS, \\ LLR, [SEE, OPTIS]}}
\rput(2.7,5.4){\rnode{PA}{\ShowX{white}{1.8cm}{Perigee \\ advance}}}
\rput(5.4,5.4){\txt{observation \\ [OPTIS]}}
\rput(4.3,3.9){\rnode{LD}{\ShowX{white}{1.8cm}{light \\ deflection}}}
\rput(7.2,3.9){\txt{VLBI \\ [Gaia, LATOR]}}
\rput(5.5,2.4){\rnode{AGR}{\ShowX{white}{1.8cm}{Absolute \\ grav. redshift}}}
\rput(7.8,2.4){\txt{GP-A \\ [OPTIS]}}
\rput(5.9,0.8){\rnode{TD}{\ShowX{white}{1.8cm}{time \\ delay}}}
\rput(8.2,0.8){\txt{Cassini}}
\rput(5.9,-0.8){\rnode{LT}{\ShowX{white}{1.8cm}{Lense-- \\ Thirring}}}
\rput(8.5,-0.8){\txt{LAGEOS, \\ [OPTIS]}}
\rput(5.5,-2.4){\rnode{SE}{\ShowX{white}{1.8cm}{Schiff \\ effect}}}
\rput(7.7,-2.4){\txt{[GP-B]}}
\rput(4.3,-3.9){\rnode{GW}{\ShowX{white}{1.8cm}{gravitational \\ waves}}}
\rput(7.4,-3.9){\txt{[laboratory \\ LISA, ASTROD]}}
\rput(2.7,-5.4){\rnode{GP}{\ShowX{white}{1.8cm}{geodetic \\ precession}}}
\rput(4.8,-5.4){\txt{LLR}}
%
%
\ncline[arrowscale=1 1]{->}{SEP}{GR}
\ncline[arrowscale=1 1]{->}{UFF}{GR}
\ncline[arrowscale=1 1]{->}{UGR}{GR}
\ncline[arrowscale=1 1]{->}{LLI}{GR}
\ncline[arrowscale=1 1]{->}{LLI1}{LLI}
\ncline[arrowscale=1 1]{->}{LLI2}{LLI}
\ncline[arrowscale=1 1]{->}{LLI3}{LLI}
\ncline[arrowscale=1 1]{<-}{PA}{GR}
\ncline[arrowscale=1 1]{<-}{LD}{GR}
\ncline[arrowscale=1 1]{<-}{AGR}{GR}
\ncline[arrowscale=1 1]{<-}{TD}{GR}
\ncline[arrowscale=1 1]{<-}{LT}{GR}
\ncline[arrowscale=1 1]{<-}{SE}{GR}
\ncline[arrowscale=1 1]{<-}{GP}{GR}
\ncline[arrowscale=1 1]{<-}{GW}{GR}
\ncline[arrowscale=1 1]{->}{NP}{GR}
\ncline[arrowscale=1 1]{->}{Gdot}{GR}
%
\end{pspicture}}
\end{center}
\caption{Tests of GR. On the left side tests the two groups of the foundations of GR are listed, on the right hand side there are predictions of GR. 
Planned missions 
are shown in brackets. We left out astrophysical observation of strong gravity effects.\label{List}}
\end{sidewaysfigure} 

\smallskip
\noindent {\bf SQUIDs.} SQUIDs are based on the flux quantization and on the Josephson effect in superconducting electrical loops. They provide the presently most sensitive magnetic flux detector. Therefore, any low--frequency signal that can be converted into a change of the magnetic flux can be observed with high precision. GP-B\cite{Everittetal01} and STEP\cite{Lockerbieetal01} use SQUIDs in order to measure the position, the linear acceleration and the angular momentum of test masses (see the talks of R. Torii and Y. Ohshima). For distance measurements the achieved sensitivity is\cite{Vodeletal01} $\delta x \sim4 \cdot 10^{-14}\;\hbox{m}/\sqrt{\hbox{Hz}}$ and for acceleration measurments $\delta a \sim 10^{-14}\;\hbox{m}/{\hbox{s}}^2/\sqrt{\hbox{Hz}}$, and for the measurement of the angular velocity one gets $\delta\omega \sim 10^{-11} \;\hbox{deg}/\hbox{s}$ for a year integration time. 

\smallskip
\noindent {\bf Cold atoms.} Using laser cooled atoms it is possible to build up highly precise and sensitive atomic interferometers. These devices can serve as highly precise accelerometers and gyroscopes, as has be demonstrated on Earth. The achieved sensitivity is $\delta a \sim 10^{-9}\;\hbox{m}/{\hbox{s}}^2/\sqrt{\hbox{Hz}}$ and $\delta\omega \sim 6 \cdot 10^{-10}\;\hbox{rad}/\hbox{s}/\sqrt{\hbox{Hz}}$. In the HYPER proposal, atomic interferometry should be used to test the UFF for quantum matter with an accuracy of $10^{-16}$ and to measure the Schiff effect (see the contribution of C. Jentsch). There are also ideas to use Bose--Einstein condensates as source for coherent sources of atoms thus enhancing the sensitivity even more. 

\smallskip
\noindent {\bf Machining.} In some cases, a high precision machining of parts of the experimental payload is uttermost important. As examples we mention the gyroscopes for GP-B and the test masses for STEP. 

\section{Summary}

In Fig.\ref{List} we list all the experiments on GR together with completed and planned missions dedicated to such tests. What is not covered by this list are searches for anomalous couplings of spin--particles with gravity, the search of a fundamental dispersion of electromagnetic propagation, and strong gravity effects like the observation of binary systems, and cosmological observations. 

\section*{Acknowledgement}

I like to thank H. Dittus, L. Iorio, and S. Scheithauer for enlightening discussions.  
Financial support from the German Space Agency DLR and the German Science Foundation DFG is greatfully acknowledged.

%
%
%
%


\begin{thebibliography}{0}

\bibitem{LaemmerzahlDittus02}
C.~L{\"a}mmerzahl and H.-J. Dittus.
\newblock {\em Ann. Physik.}, 11:95, 2002.

\bibitem{GiuliniKieferLaemmerzahl03}
C.~Kiefer, D.~Giulini, and C.~L{\"a}mmerzahl.
\newblock {\em Quantum Gravity -- From Theory to Experimental Search}.
\newblock Springer--Verlag, Berlin, 2003.

\bibitem{Ni77}
W.-T. Ni.
\newblock {\em Phys.\ Rev.\ Lett.}, 38:301, 1977.

\bibitem{Will93}
C.M. Will.
\newblock {\em Theory and Experiment in Gravitational Physics (Revised
  Edition)}.
\newblock Cambridge University Press, Cambridge, 1993.

\bibitem{DamourPiazzaVeneziano02}
T.~Damour, F.~Piazza, and G.~Veneziano.
\newblock {\em Phys. Rev. Lett.}, 89:081601, 2002.

\bibitem{DamourPiazzaVeneziano02a}
T.~Damour, F.~Piazza, and G.~Veneziano.
\newblock {\em Phys. Rev.} {\bf D 66}, 046007 (2002).

\bibitem{Carrolletal01}
S.M. Carroll {\it et al}.
\newblock {\em Phys. Rev. Lett.} {\bf 87}, 141601 (2001).

\bibitem{Antoniadis03}
I.~Antoniadis.
\newblock Physics with large extra dimensions and non--newtonian gravity at
  sub--mm distances.
\newblock In [2]. 

\bibitem{Ciufolinietal98}
I.~Ciufolini {\it et al}.
\newblock {\em Science} {\bf 279}, 2100 (1998).

\bibitem{IorioMorea04}
L.~Iorio and A.~Morea.
\newblock The impact of the new {E}arth gravity models on the measurement of
  the {L}ense--{T}hirring effect.
\newblock {\em Gen. Rel. Grav.}, 36:to appear, 2004, [gr-qc/0304011].

\bibitem{Iorioetal03}
L.~Iorio {\it et al}.
\newblock On the possibility of measuring the {L}ense--{T}hirring effect with a
  {LAGEOS--LAGEOS II--OPTIS}--mission, 2003.
\newblock preprint gr-qc/0211013.

\bibitem{Laemmerzahletal04}
C.~L{\"a}mmerzahl {\it et al}.
\newblock {\em Gen. Rel. Grav.}, 36:615, 2004.

\bibitem{Everittetal01}
C.W.F. Everitt {\it et al}.
\newblock In C.~L{\"a}mmerzahl, C.W.F. Everitt, and F.W. Hehl, editors, {\em
  Gyros, Clocks, and Interferometers: Testing Relativistic Gravity in Space},
  page~52. Springer--Verlag, Berlin, 2001.

\bibitem{Salomonetal01}
C.~Salomon {\it et al}.
\newblock {\em C.R. Acad. Sci. Paris} {\bf 4}, 1313 (2004).

\bibitem{Touboul01a}
P.~Touboul.
\newblock {\em Comptes Rendus de l'Aced. Sci. S\'erie IV: Physique
  Astrophysique} {\bf 2}, 1271 (2001).

\bibitem{Lockerbieetal01}
N.~Lockerbie, J.C. Mester, R.~Torii, S.~Vitale, and P.W. Worden.
\newblock In C.~L{\"a}mmerzahl, C.W.F. Everitt, and F.W. Hehl, editors, {\em
  Gyros, Clocks, and Interferometers: Testing Relativistic Gravity in Space},
  page 213. Springer--Verlag, Berlin, 2001.

\bibitem{Nietal02}
W.-T. Ni~et al.
\newblock {\em Int. J. Mod. Phys.}, D 11:1035, 2002.

\bibitem{TuryshevShaoNordtvedt03}
S.G. Turyshev, M.~Shao, and K.~Nordtvedt.
\newblock New concept for testing general relativity: The laser astrometric
  test of relativity (lator) mission, 2003.
\newblock gr-qc/0311049.

\bibitem{TuryshevShaoNordtvedt04}
S.G. Turyshev, M.~Shao, and K.~Nordtvedt.
\newblock The laser astrometric test of relativity mission, 2004.
\newblock gr-qc/0401063.

\bibitem{Laemmerzahletal04a}
C.~L{\"a}mmerzahl {\it et al}.
\newblock An {E}instein mission for improved tests of special and general
  relativity, 2004.
\newblock preprint.

\bibitem{MalekiPrestage01}
L.~Maleki and J.~Prestage.
\newblock In C.~L{\"a}mmerzahl, C.W.F. Everitt, and F.W. Hehl, editors, {\em
  Gyros, Clocks, and Interferometers: Testing Relativistic Gravity in Space},
  page 369. Springer--Verlag, Berlin, 2001.

\bibitem{WEbpage}
See the informations available on {\tt http://tycho.dm.unipi.it/~nobili/}

\bibitem{Sandersetal00}
A.J. Sanders {\it et al.}
\newblock {\em Class. Quantum Grav.}, 17:2331, 2000.

\bibitem{Touboul01}
P.~Touboul.
\newblock In C.~L{\"a}mmerzahl, C.W.F. Everitt, and F.W. Hehl, editors, {\em
  Gyroscopes, Clock, Interferometers, ...: Testing Relativistic Gravity in
  Space}, volume LNP 562, page 274. Springer, Berlin, 2001.

\bibitem{Prestageetal01}
J.D. Prestage, S.~Chung, E.~Burt, L.~Maleki, and R.L. Tjoelker.
\newblock Proceedings of the 2001 IEEE International Frequency Control
  Symposium.

\bibitem{Muelleretal03c}
H.~M{\"u}ller, S.~Herrmann, C.~Braxmaier, S.~Schiller, and A.~Peters.
\newblock {\em Phys. Rev. Lett.}, 91:020401, 2003.

\bibitem{Wolfetal04}
P.~Wolf, M.E. Tobar, S.~Bize, A.~Clairon, A.N. Luiten, and G.~Santarelli.
\newblock {\em Gen. Rel. Grav.} {\bf 36}, to appear 2004.

\bibitem{CundiffYeHall01}
S.T. Cundiff, J.~Ye, and J.L. Hall.
\newblock {\em Rev. Sci. Instr.} {\bf 72}, 3749 (2001).

\bibitem{Vodeletal01}
W.~Vodel {\it et al}.
\newblock In C.~L{\"a}mmerzahl, C.W.F. Everitt, and F.W. Hehl, editors, {\em
  Gyroscopes, Clock, Interferometers, ...: Testing Relativistic Gravity in
  Space}, volume LNP 562, page 248. Springer, Berlin, 2001.

\end{thebibliography}
\end{document}